\documentclass[reqno,11pt]{article}
\usepackage{ifpdf}
\usepackage[usenames,dvipsnames]{xcolor}
\usepackage{ae}
\usepackage[T1]{fontenc}
\usepackage[ansinew]{inputenc}
\usepackage{mathrsfs}
\usepackage{amsmath}
\usepackage{amssymb}
\usepackage{graphicx}
\usepackage{color}
\definecolor{darkblue}{cmyk}{0.9,0.9,0,0}
\definecolor{darkgreen}{rgb}{0,0.55,0}
\usepackage[colorlinks=true,linkcolor=darkblue,citecolor=darkblue,urlcolor=darkblue]{hyperref}
\usepackage{epsfig}

\newcommand{\comment}[1]{}

\newcommand{\cG}{{G}}
\newcommand{\beq}{\begin{equation}}
\newcommand{\eeq}{\end{equation}}
\newcommand{\beqq}{\begin{equation*}}
\newcommand{\eeqq}{\end{equation*}}
\newcommand\beqa{\begin{eqnarray}}
\newcommand\eeqa{\end{eqnarray}}
\newcommand\beqaa{\begin{eqnarray*}}
\newcommand\eeqaa{\end{eqnarray*}}
\newcommand\bea{\begin{array}}
\newcommand\eea{\end{array}}

\def\XXint#1#2#3{{\setbox0=\hbox{$#1{#2#3}{\int}$}
\vcenter{\hbox{$#2#3$}}\kern-.5\wd0}}

\newcommand{\neqa}{\nonumber\end{eqnarray}}
\newcommand{\la}[1]{\label{#1}}

\newcommand{\eq}[1]{(\ref{#1})}

\def\tr{{\rm tr~}}

\renewcommand{\d}{\partial}

\newcommand{\<}{{\langle}}
\renewcommand{\>}{{\rangle}}

\newcommand{\re}{\relax{\rm I\kern-.18em R}}

\renewcommand{\sp}{p\hspace{-.40em}/}

\def\su2{{SU(2)}}

\def\[{\left[}
\def\]{\right]}

\def\({\left(}
\def\){\right)}
\def\[{\left[}
\def\]{\right]}

\def\<{\langle}
\def\>{\rangle}

\def\i2{\frac{i}{2}}

\def\spi{\relax{\rm \pi\kern-0.5em /}}
\def\sA{\relax{\rm A\kern-0.5em /}}
\def\sp{\relax{\rm p\kern-0.5em /}}
\def\sd{\relax{\rm \d\kern-0.5em /}}
\def\sk{\relax{\rm k\kern-0.5em /}}
\def\sn{\relax{\rm n\kern-0.5em /}}
\def\sl{\relax{\rm l\kern-0.5em /}}
\def\sP{\relax{\rm P\kern-0.7em /}}
\def\sBethe{\relax{\rm \Bethe\kern-0.5em /}}

\usepackage{varioref}
\usepackage{makeidx}
\makeindex

\usepackage[english]{babel}
\begin{document}

\renewcommand{\thefootnote}{\fnsymbol{footnote}}
\setcounter{page}{1}
\setcounter{footnote}{0}
\setcounter{figure}{0}
\begin{center}
{\Large\textbf{\mathversion{bold}
On the Derivation of the Exact Slope Function
}\par}

\vspace{0.5cm}

\textrm{Nikolay Gromov}
\\
\footnotesize{
\textit{King's College London, Department of Mathematics WC2R 2LS, UK \& \\
St.Petersburg INP, St.Petersburg, Russia} \\
\texttt{nikgromov@gmail.com}
\vspace{3mm}
}

\par\vspace{0.7cm}

\textbf{Abstract}
\end{center}
In this note we give a simple derivation of the exact slope function conjectured by Basso
for the anomalous dimensions
of Wilson operators in the ${\frak s\frak l}(2)$ sector of planar
${\cal N}=4$ Super-Yang-Mills theory.
We also discuss generalizations of this result
for higher charges and other sectors.
\noindent
\section{Introduction}
It was discovered by Basso \cite{Basso} that the smallest anomalous dimension $\gamma$
of the ${\frak s\frak l}(2)$ operators $\tr D^S Z^J$ in the planar ${\cal N}=4$ Super-Yang-Mills theory in the limit of
small spin
is given by the universal function
\beq\la{slope}
\gamma = S\frac{\Lambda }{J} \frac{I_{J+1}(\Lambda)}{I_J(\Lambda)}+{\cal O}(S^2)\;
\eeq
where $I_J$ is the modified Bessel function of the first kind, $\Lambda=n\sqrt{\lambda}$, $\lambda$ is the `t Hooft coupling,
$n$ is the integer mode number characterizing the state,
$J$ is the twist and $S$ the Lorentz spin.
The simplicity of \eq{slope} is partially due to the assumption that
wrapping corrections are suppressed in this
limit and one can use the asymptotic Bethe ansatz (ABA). This assumption,
leading to enormous simplifications, was tested both at weak and strong coupling
and still remains mysterious to a large extent. We will also neglect wrapping effects
and use the ABA as the starting point.

There are various applications of the result \eq{slope}. First of all, already in \cite{Basso}
it was proposed that this equation can be used together with some natural assumptions
about the structure of the strong coupling expansion to confirm the numerical prediction
of the Y-system and TBA equations for the one loop coefficient of the Konishi state ($J=2,S=2$).
Furthermore, in \cite{Basso} it was proposed that it can be used to predict the next
two loop term in the strong coupling expansion of the Konishi dimension. This coefficient was found
in \cite{Saulius}
in agreement with the TBA numerics for various Konishi--like states \cite{GKV2}.
Recently it was proposed \cite{Sever} that the exact slope function \eq{slope} may become a bridge between
integrability and localization techniques which were completely detached from each other previously.
In any case the result \eq{slope} is very exciting and still contains many puzzles.

Initially \eq{slope} was derived in \cite{BassoToAppear}
for a particular case $J=2$ and $n=1$ using the Baxter equation.
Here we develop a different approach which applies for general situation\footnote{See the note added.}.

We derive \eq{slope} in two steps. First, we assume $\Lambda$-scaling of the slope function which, as can be seen from \eq{slope}, depends on a particular
combination of the `t Hooft coupling and the mode number. This allows one to take the limit $n\to 0$, simplifying ABA.
Second, we show that the obtained solution is independent of the mode number.

As a generalization of \eq{slope} we also found a simple expression for the slope function of the higher charges
$
{\cal Q}_{r+1} = S\frac{\Lambda }{J} \frac{I_{J+r}(\Lambda)}{I_J(\Lambda)}+{\cal O}(S^2)
$,
as well as its analog in the $\frak{su}(2)$ sector.

 \def\nref#1{{(\ref{#1})}}

\section{Derivation of the Exact Slope Function}
As we announced in the introduction our starting point is the asymptotic Bethe ansatz
\cite{AFS}
\beq\la{BAE}
\frac{J}{i}\log\(\frac{x_k^+}{x_k^-}\)
-\sum_{j\neq k}^S\frac{1}{i}\log\(\frac{x_k^--x_j^+}{x_k^+-x_j^-}\frac{1-1/(x_k^+x_j^-)}{1-1/(x_k^-x_j^+)}
\sigma^2(u_k,u_j)\)
=2\pi n_k\;\;,\;\;k=1,\dots,S
\eeq
where $x_k^\pm\equiv x(u_k\pm i/2)$, $x_k\equiv x(u_k)$  with $x(u)=2\pi u/\sqrt\lambda+\sqrt{4\pi^2u^2/\lambda-1}$. The function $\sigma^2(u_k,u_j)$
is the dressing phase \cite{BES}. For our derivation we
need only its leading order strong coupling expression \cite{AFS} as we explain below:
\beq
\frac{2}{i}\log\sigma(u_k,u_j)\simeq-\frac{2}{i}\log\(\frac{1-1/(x_k^+ x_j^-)}{1-1/(x_k^- x_j^+)}\)
+2(u_j-u_k)\log\(\frac{x_j^- x_k^--1}{x_j^- x_k^+-1}
\frac{x_j^+ x_k^+-1}{x_j^+ x_k^--1}\)\;.
\eeq
The scaling dimension $\Delta$ and its anomalous part $\gamma$ are given in terms of the Bethe roots
\beq
\Delta=J+S+\gamma\;\;,\;\;\gamma=\frac{i\sqrt\lambda}{2\pi}\sum_{j=1}^S \(\frac{1}{x_j^+}-\frac{1}{x_j^-}\)\;.
\eeq
The integer numbers $n_k$ in the r.h.s. of \eq{BAE} are the mode numbers distinguishing different states
of the ${\frak sl}(2)$ sector. First we assume all of them to be the same $n_k=n$\footnote{
This assumption does not reduce the generality.
Roots with different $n_k$ in the limit of vanishing number of roots do not interact with each
other (see footnote [25] in \cite{Basso}). As a consequence the result for the anomalous dimension is expected to be
$\gamma~=~\sum\limits_{k=1}^S \frac{n_k\sqrt\lambda}{J} \frac{I_{J+1}(n_k\sqrt\lambda)}{I_J(n_k\sqrt\lambda)}+{\cal O}(S^2)$.
}.
The main trick in our derivation is to take the $n\to 0$ limit.
Since it is expected that the result only depends on the combination $\Lambda=n\sqrt\lambda$
we do not lose any information. Yet the Bethe equations simplify considerably in this limit.
First of all if $\Lambda$ is fixed we see that $\lambda\sim 1/n^2\to\infty$ which ensures that only
the leading strong coupling piece of the dressing phase contributes\footnote{we will argue that the result does not depend on the dressing phase at all.}! In this limit one also has $u_k\sim 1/n$.
Expanding \eq{BAE} we get simply
\beq\la{exp}
\sum_{j\neq k}
\frac{2}{x_k-x_j}+\frac{1}{x_k}\(J+\gamma+\frac{2}{1-x_k^2}\)=
\frac{\Lambda(x_k^2-1)}{2 x_k^2}\;.
\eeq
The anomalous dimension which can be written in terms of the resolvent
\beq\la{defRes}
{G}(x)=\sum_{j=1}^S\frac{1}{x-x_j}
\eeq
becomes
\beq\la{energy}
\gamma=-{G}(+1)+{G}(-1)\;.
\eeq

The equation \eq{exp} is an equation typically appearing in matrix models.
We use the standard trick  of multiplying \eq{exp} by $\frac{1}{x-x_k}$,
summing over $k$ and using exact identities like
$
\sum_{j\neq k}\frac{2}{(x-x_k)(x_k-x_j)}={G}^2(x)+{G}'(x)
$
in the end we get
\beq\la{exact}
{G}^2(x)+{G}'(x)
+\(\frac{J+\gamma+2}{x}
-\frac{2x}{x^2-1}
+\frac{\Lambda}{2}\frac{1-x^2}{x^2}\)G(x)
=F(x)
\eeq
where
\beq\la{exact2}
F(x)=
(J+\gamma+2)\frac{\cG(0)}{x}
+\frac{\Lambda}{2}\frac{\cG(0)+\cG'(0)x}{x^2}
-\frac{\cG(+1)}{x-1}
-\frac{\cG(-1)}{x+1}
\;.
\eeq
Note that in the large $x$ limit $G(x)\sim S/x$ and \eq{exact} gives:
\beq
\Lambda {G}'(0)=2\cG(+1)+2\cG(-1)-2{G}(0)(J+\gamma+2)-\Lambda S\;
\eeq
which allows us to get rid of ${G}'(0)$ in \eq{exact2} and explicitly introduce the parameter $S$
which now can be an arbitrary number (non necessarily an integer):
\beq\la{FL}
F(x)=
\frac{G(0)}{2x^2}+\frac{2G(+1)+2G(-1)-\Lambda S}{2x}
-\frac{G(+1)}{x-1}
-\frac{G(-1)}{x+1}\;.
\eeq

So far we have not used that $S$ is small.
Note that $G(x)\sim S$ as it is defined as a sum of $S$ terms.
This implies that to leading order in $S$ we can drop the terms $G^2(x),\;G(x)\gamma$ and $G(0)\gamma$.
After that \eq{exact} becomes a first order linear differential equation which
can be immediately integrated:
\beq\la{Gint}
{G}(x)=\frac{x^2-1}{x^{J+2}}e^{\Lambda\frac{x^2+1}{2x}}
\int_{x_0}^x
F(y)
\frac{y^{J+2}}{y^2-1}e^{-\Lambda\frac{y^2+1}{2y}}dy\;.
\eeq
$x_0$ is a constant of integration.
It should be fixed by requiring analyticity of $G(x)$.
In order for $G(x)$ to be finite at the origin we must set $x_0=0$.
After that we find the resolvent as a function of three yet to be fixed  parameters $G(0)$
and $G(\pm 1)$. They can also be found by requiring further analyticity.
The integrand has poles at $y=\pm 1$ which may lead
to a logarithmic singularity at $x=\pm 1$ if the residues are not zero. Requiring the residues
at $y=\pm 1$ to vanish we get
\beqa
\Lambda (G(0)-S)-G(+1)(1+2J)+G(-1)=0\quad,\quad
\Lambda (G(0)+S)+G(-1)(1+2J)-G(+1)=0\;.
\eeqa
This further simplifies $G(x)$. Integrating by parts we get explicitly
\beq\la{Glast}
{G}(x)=-\frac{\Lambda S}{2J}-\frac{\gamma}{2x}-
\frac{x^2-1}{x^{J+2}}e^{\Lambda\frac{x^2+1}{2x}}\frac{\Lambda}{4J}
\int_{0}^x
{dy}\left(\gamma J y^{J-1}+\Lambda S y^J\right)
e^{-\Lambda\frac{y^2+1}{2y}}\;,
\eeq
which still contains the unknown anomalous dimension $\gamma$. We find it by requiring
analyticity at the origin. Indeed for an arbitrary $\gamma$ the origin is a branch point.
The integral has a nontrivial monodromy
which can be evaluated in terms of the modified Bessel functions
\beq
I_\nu(\Lambda)=\frac{(-1)^{-\nu}}{2\pi i}\oint
y^{\nu-1}e^{-\Lambda\frac{y^2+1}{2y}}\;
\eeq
where the integration contour starts at the origin, goes around counterclockwise
and returns back to the origin. We see that in order for the integral \eq{Glast} to be regular
the two terms in the integrand should be tuned in a precise way:
\beq\la{derslope}
\gamma J (-1)^JI_J(\Lambda)+\Lambda S(-1)^{J+1} {I_{J+1}(\Lambda)}=0\;
\eeq
which indeed leads to \eq{slope}.

\subsection{Exact Slope of Higher Local Charges}
The anomalous dimension $\gamma$
is only one representative of an infinite family of local conserved charges.
Their eigenvalues are given in terms of the Bethe roots by
\cite{AFS}
\beq\la{localq}
{\cal Q}_r\equiv \frac{\sqrt\lambda}{2\pi}\sum_{j=1}^S\(\frac{i (x_j^+)^{1-r}}{r-1}-\frac{i (x_j^-)^{1-r}}{r-1}\)\;.
\eeq
The first $r=1$ charge can be computed exactly by summing up all Bethe equations \eq{BAE}:
\beq\la{exactQ1}
{\cal Q}_1=\sum_j\frac{\Lambda}{2\pi n}\frac{1}{i}\log\frac{x_j^+}{x_j^-}=\frac{\Lambda S}{J}\;,
\eeq
the anomalous dimension coincides in our normalization with the second charge $\gamma={\cal Q}_2$.
The other charges are not directly related to the global symmetry
generators. Nevertheless, they may also be of some interest.
In this section we extend the result of the previous section and find an explicit expression for all the charges.
To compute them we introduce their generating function
\beq\la{Hdef}
H(x)\equiv\sum_{j=1}^S\frac{\Lambda}{4\pi i n}\log\frac{x/x^-_j-1}{x/x^+_j-1}
=-\frac{1}{2}\sum_{r=1}{\cal Q}_{r+1} x^{r}\;.
\eeq
As before we assume that it is enough to consider the $n\to 0$ limit.
We give an argument in the next section that the results we obtain in this limit
are correct for any $n$.
In this limit we can write $H(x)$ in terms of the resolvent $G(x)$:
\beq\la{Gint}
H(x)\to H_0(x)\equiv
\sum_{k=1}^S \frac{1}{x-x_j}\frac{x x_j}{x_j^2-1}
=\frac{x^2 G(x)}{x^2-1}
+\frac{\Lambda  S x^2+\gamma J x}{2 J \left(x^2-1\right)}\;\;,\;\;n\to 0
\eeq
using \eq{Glast} and \eq{derslope} we have
\beq\la{H0}
H_0(x)=-
\frac{e^{\Lambda\frac{x^2+1}{2x}}}{x^{J}}\frac{S\Lambda^2}{4J}
\int_{0}^{x}
{dy}\left(\frac{I_{J+1}(\Lambda)}{I_J(\Lambda)} y^{J-1}+y^J\right)
e^{-\Lambda\frac{y^2+1}{2y}}+{\cal O}(S^2)\;.
\eeq
Expansion in small $x$ of this integral is straightforward. What we find is a very simple expression
\beq\la{H0exp}
H_0(x)=-\frac{\Lambda S}{2J}\sum_{r=1}^\infty\frac{I_{J+r}(\Lambda)}{I_J(\Lambda)}x^r
\eeq
which for the charges gives
\beq\la{Qrs}
{\cal Q}_r=S\frac{\Lambda}{J}\frac{I_{J+r-1}(\Lambda)}{I_J(\Lambda)}+{\cal O}(S^2)\;.
\eeq
Note that \eq{H0exp} has infinite radius of convergence. It is also clear from the integral representation
\eq{H0}
that $H_0(x)$ is an entire function.\footnote{It may seems a little surprising that the resolvent
is regular everywhere on the complex plane. Normally resolvents are singular at the Bethe roots
and the singularity may stay even when the number of roots tends to zero. The explanation of this paradox
is that the singularity corresponding to the location of the roots
moves to infinity as one decreases $S$. This can be seen, for example, by solving
the exact nonlinear differential equation \eq{exact}.
}
The simplicity of the result \eq{H0exp} implies the following nice identities for $H_0(x)$
\beq\la{infinity}
x^JH_0(x)=P_J(u)-\frac{1}{x^J}H_0(1/x)-\frac{\Lambda S }{2J I_J(\Lambda)}e^{2\pi n u}\;
\eeq
where $P_J(u)$ is a polynomial of degree $J$ given by $P_J(u)=-\frac{\Lambda S}{4J}\sum_{r=-J}^J\frac{I_{r}(\Lambda)}{I_J(\Lambda)}\(x^r+\frac{1}{x^r}\)$.
This identity immediately follows from $e^{2\pi n u}=e^{\Lambda \frac{x^2+1}{2x}}=\sum_{r=-\infty}^\infty I_r(\Lambda) x^r$.
Note that the first two terms are regular at infinity
whereas the last term reveals an essential singularity of $H_0(u)$ at infinity.
We will use these analytical properties in the next section for the proof of $\Lambda$-scaling.

\subsection{Proof of $\Lambda$-Scaling}
In the previous section our consideration was limited to the case of small mode number $n$
and fixed coupling $\Lambda=n\sqrt\lambda$. It was conjectured in \cite{Basso}
that the exact slope function only depends on this combination of `t Hooft coupling and the mode number $n$.
In this section we prove, on a physical level of rigor, a bit more general statement, namely that all local charges obey this property.
In other words, we show that the generating function of the local charges $H(x)$ does not depend on $n$
and thus coincides with its small $n$ limit $H_0(x)$ given by \eq{H0exp} for small $S$.

To show this we make several important observations. Firstly,
one can add or subtract from the Bethe ansatz equations \eq{BAE}
the terms which are not singular when $u_k-u_j\sim 1$.
Indeed, such terms are clearly of order $S$ and are irrelevant when $S\to 0$.
In particular the whole dressing phase $\sigma^2(u_k,u_j)$, which is not singular when $u_k\sim u_j$,
is of no relevance for our considerations. It would be convenient
to replace \eq{BAE} by
\beq\la{uglyBAE}
e^{-2\pi i n}\(\frac{x_k^+}{x_k^-}\)^{J}\prod_{j\neq k}^S
\frac{x_k^+-x_j^-}{x_k^--x_j^+}
\prod_{j=1}^S\frac{x_j^+}{x_j^-}\frac{1/x_k^+-x_j^+}{1/x_k^--x_j^-}=1\;\;,\;\;k=1,\dots,S
\eeq
which to first order in $S$ should give the same result as \eq{BAE}.
Secondly, \eq{uglyBAE} can be understood as a condition of pole cancelation
of a Baxter-like function $T(u)$ defined by
\beq\la{Tdef}
T(u)\equiv e^{-i\pi n}{(x^+)^{J}}\prod_{j=1}^S\frac{x^+/x_j^--1}{x^+/x_j^+-1}+
e^{+i\pi n}{(x^-)^{J}}\prod_{j=1}^S\frac{x^-/x_j^+-1}{x^-/x_j^--1}\;.
\eeq
Indeed, $T(u)$ may have poles at $u=u_k$. When \eq{uglyBAE} is satisfied these poles
cancel between the first and the second terms in \eq{Tdef}.
The only singularities which $T(u)$ may have are the branch points at $u=2g\pm i/2$ originating
from $x^\pm = x(u\pm i/2)$.

The Baxter equation \eq{Tdef} can be understood as an equation on $H(x)$
\beq\la{bax}
T(u)\equiv e^{-i\pi n}{(x^+)^{J}}e^{\frac{4\pi i n}{\Lambda} H(x^+)}+
e^{+i\pi n}{(x^-)^{J}}e^{\frac{4\pi i n}{\Lambda} H(x^-)}\;
\eeq
where $T(u)$ should be found self-consistently by the requirement of
analyticity. In particular at infinity
$T(u)$ should be regular and behave as
\beq\la{largexofT}
T(u)= 2x^J \cos(\pi n-2\pi n S/J)+{\cal O}(x^{J-1})
\eeq
as one can see from \eq{Tdef} and \eq{exactQ1}.

Let us show that for any mode number $n$
$H=H_0$
solves the Baxter equation \eq{bax} when $S\to 0$. Using that $H\sim S$ we get
\beq\la{Tex}
T(u)=e^{-i\pi n}{(x^+)^{J}}\(1+\frac{4\pi i n}{\Lambda}H(x^+)\)+
e^{i\pi n}{(x^-)^{J}}\(1-\frac{4\pi i n}{\Lambda}H(x^-)\)+{\cal O}(S^2)\;.
\eeq
The function $H_0(u)$ defined by \eq{H0exp} is an entire function
with an essential singularity at infinity. This means that in principle
$T(u)$ can be also singular at infinity. If this is the case
our guess that $H(u)=H_0(u)$ for any mode number $n$ must be wrong.
Fortunately, the singularity at infinity does cancel!
Indeed, from \eq{infinity} we can write
\beq
x^J H_0(x)=-\frac{\Lambda S }{2J I_J(\Lambda)}e^{2\pi n u}+\(-\frac{\Lambda S}{2J}x^J+O(x^{J-1})\)\;,
\eeq
where only the first term is singular
and in the combination \eq{Tex} the singular terms cancel  and the large $u$ asymptotic
of $T(u)$ is
\beq\
T(u)= e^{-i\pi n}x^J\(1-\frac{2\pi i n S}{J}+{\cal O}(S^2)\)+c.c.+{\cal O}(x^{J-1})
\eeq
which coincides with \eq{largexofT}!

The above arguments show that $H_0$ is a solution of the Baxter equation.
To make our proof more rigorous we have to argue for the uniqueness of the construction.
First, the solution  of the Baxter equation is not unique.
There are many various configurations of Bethe roots each corresponding to some
solution of the Baxter equation. As we discussed in the beginning these
configurations differ by their mode numbers $n_k$ which for the twisted Bethe equation
\eq{uglyBAE} could take values $n+k$ for some integers $k$.
However, at the level of the Baxter equation there are no mode numbers
and we have to find a definition for them. For that we notice
that the allowed behavior of $H$ at infinity is very restricted by \eq{largexofT}
to
\beq
x^J H= \sum_k A_k e^{2\pi (n+k) u}+x^J\times  {\rm regular\;at\;infinity}\;
\eeq
since the first term drops out from $T(u)$. It is of course natural to define
the single cut configuration by saying that all $A_k$ except one are zero.
The general configuration is given by a linear combination of these solutions.
Thus we restrict ourselves to the solution with $k=0$
\beq\la{Hmy}
x^J H(x)= A_0 e^{\Lambda \frac{x^2+1}{2x}}+x^J\times{\rm regular\;at\;infinity}\;
\eeq
we used that $2\pi n u=\Lambda \frac{x^2+1}{2x}$.
Furthermore the function $H(x)$ should be an entire function.
Indeed if $H$ was singular at some point $u=u_0$
then from \eq{Tex} it would be also singular at $u_0+i m,\;m\in{\mathbb Z}$.
These singularities would accumulate at $x(u)=\infty$ in contradiction with \eq{Hmy}
or at $x(u)=0$ in contradiction with the definition \eq{Hdef}.
Since $T(u)$ grows as $x^J$
at infinity the regular at infinity part of $H(x)$ should
not grow faster than $x^0$.
We can get rid of the unknown second term in \eq{Hmy} by multiplying
it by $\frac{1}{x-y}-\frac{1}{x}$ and integrating around infinity:
\beq
\frac{1}{2\pi i}\oint  H(x) \[\frac{1}{x-y}-\frac{1}{x}\]dx  =
\sum_{r=1}^\infty\frac{y^{r}}{2\pi i}
\oint A_0 e^{\Lambda \frac{x^2+1}{2x}}\frac{1}{x^{J+r+1}}  {dx}=
\sum_{r=1}^\infty A_0 y^r I_{J+r}(\Lambda)\;.
\eeq
On the other hand contracting the contour to the origin in the l.h.s.
we get $H(y)$ since $H(0)=0$ due to \eq{Hdef}. Finally comparing
with \eq{largexofT} we fix $A_0=-\frac{\Lambda S}{2 J I_J}$
and indeed reproduce \eq{H0exp}.

\section{Exact Slope Function for the ${\mathfrak s \mathfrak u}(2)$ Sector}
One can also attempt to define the exact slope function
for operators in ${\frak su}(2)$ sector of the type $\tr(X^M Z^{L-M})$.
The role of the small spin $S$ would be played by $M$. An obvious problem here is that
$M$ cannot exceed $L$. In order to be able to extrapolate the anomalous dimension
to zero $M$ one should let $L$ be arbitrarily large. Despite these potential problems
the derivation from the previous sections can be literary repeated for this case.
One simply finds that in \eq{Gint} and \eq{FL} $J$ gets replaced by $-L$ and $\Lambda$ by $-\Lambda$.
This results in a simple replacement of the parameters in the ${\frak sl}(2)$ slope function
\beq\la{slopesu2}
\gamma = -M\frac{\Lambda }{L} \frac{I_{-L+1}(\Lambda)}{I_{-L}(\Lambda)}+{\cal O}(M^2)\;.
\eeq
This expression needs to be dealt with some care as we now explain.
We notice that for a positive integer $L$ the expansion of $\gamma$ starts from $\Lambda^0$ and not from $\Lambda^2$
as it should be for the anomalous dimension $\gamma$ which must vanish at tree level.
One of the ways to make sense of \eq{slopesu2} is to first expand in small $\Lambda$
and only then set $L$ to its integer value. This prescription gives
a finite result provided $L$ is bigger then the order in perturbation theory. For example the first couple of orders
are
\beqa\la{Expansion}
\frac{\gamma}{M}&\simeq&\frac{\Lambda ^2}{2 (L-1) L}-\frac{\Lambda ^4}{8 (L-2) (L-1)^2 L}
+\frac{\Lambda ^6}{16 (L-3)
   (L-2) (L-1)^3 L}
   +{\cal O}(\Lambda^{8})
\eeqa
we see that at the order $\Lambda^{2l}$ the expansion coefficient has poles at $L=0,1,\dots,l$.
A similar prescription should be also applied to the higher charges.
We verified numerically that the above procedure gives the correct result with the relative error $\sim 10^{-46}$
up to $\Lambda^8$ for $L=300,400,500$\footnote{We solve the ABA for a small number of roots $M=1,\dots,30$ and then fit the
 anomalous dimension  by a polynomial.
The linear coefficient of this polynomial was then compared to the analytical prediction \eq{Expansion}.
This procedure gives enough precision to confidently test all the $1/L$ terms in \eq{Expansion}, but
it hardly can exclude exponentially small terms which may be missing in \eq{Expansion}.
}.

Note that since the cuts do not interact with each other and any cut can be brought to a rank one sector by
a suitable duality transformation our results cover the most general case.

\subsection{Derivation of the functional expression}
\beq\la{Gint2}
{G}(x)=\frac{x^2-1}{x^{-J+2}}e^{-\Lambda\frac{x^2+1}{2x}}
\int_{x_0}^x
F(y)
\frac{y^{-J+2}}{y^2-1}e^{+\Lambda\frac{y^2+1}{2y}}dy\;.
\eeq
This time we have to set $x_0=-\infty$ since then there is a hope to cancel
the divergence at $-\infty$ from the prefactor. After that we can again integrate by parts to get
\beq\la{Glast2}
{G}(x)=-\frac{\Lambda S}{2J}-\frac{\gamma}{2x}+
\frac{x^2-1}{x^{-J+2}}e^{-\Lambda\frac{x^2+1}{2x}}\frac{\Lambda}{4J}
\int_{-\infty}^x
{dy}\left(\gamma J y^{-J-1}+\Lambda S y^{-J}\right)
e^{+\Lambda\frac{y^2+1}{2y}}\;,
\eeq

\section{Discussion}
Our derivation of the exact slope function is very universal and can be applied with almost
no changes to numerous theories where spectrum is described by a Bethe ansatz.
In particular we can see that for the ABJM theory one should find \cite{Beccaria:2012qd} the same result
when written in terms of a yet unknown interpolation function $h(\lambda)$.
For that to be true one should assume that wrapping corrections
are negligible.
The wrapping corrections indeed seem to be suppressed when $S\to 0$ in the ${\cal N}=4$ SYM
but in ABJM theory the slope function of \cite{Basso} gets non-zero corrections from wrapping \cite{Beccaria:2012qd}.
Still it should describe all contributions of the type $1/J^n$.
It should be possible to compute these wrapping corrections using the Y-system approach.
This may be interesting since in \cite{Sever} it was argued that this information
may help fixing the effective coupling $h(\lambda)$ in terms of the `t Hooft coupling $\lambda$.

Another comment we would like to make is also related to \cite{Sever}.
They found that the exact expression for the Bremsstrahlung function in ${\cal N}=4$ SYM is given by
$
B=\frac{1}{4\pi^2}\lambda\d_\lambda \log\langle W_\circledcirc \rangle
$
where $\langle W_\circledcirc \rangle$ is the 1/2 BPS circular Wilson loop vacuum expectation value. It
was computed for {\it finite} $N$ for the $U(N)$ gauge group in
\cite{Erickson:2000af} to be $\langle W_\circledcirc \rangle= \frac{1}{N}L_{N-1}^1(-\lambda/4N)e^{\frac{\lambda}{8N}}$\footnote{$L$ is the modified Laguerre polynomial}.
What is exciting is that in the planar limit $N\to\infty$ it gives
$
B=\frac{1}{4\pi^2}\frac{\sqrt\lambda I_2(\sqrt\lambda)}{I_1(\sqrt\lambda)}
$
which is nothing but \eq{slope} for $J=1$ up to a simple factor. In view of this nice observation it is very
appealing to say that a similar expression may give the non-planar generalization of the exact slope function.
For example simply replacing $L_{N-1}^1(-\lambda/4N)$ by $L_{N-1}^J(-\lambda/4N)$ would give the correct result in the planar limit.
It would be interesting to investigate this question.
It seems that the small $S$ limit would be a natural direction to attack non-planarity
in the spectral problem
since in this limit one should find many simplifications and may even hope to get exact analytical results.

{\bf Note added} When this note was ready we became aware that the initial derivation for $J=2$ of \cite{BassoToAppear} was extended to
arbitrary $J$ and $n$ and will appear in \cite{BassoToAppear}. We are grateful to Benjamin Basso for sharing these results before the publication.

{\bf Acknowledgments}
We thank Saulius Valatka for collaboration on the initial stage of the project, Benjamin Basso, Vladimir Kazakov, Pedro Vieira,
 Amit Sever and Konstantin Zarembo for discussions and the
 Israel IAS of The Hebrew University of Jerusalem
 for the kind hospitality.
\hspace{-10mm}

\end{document}